\documentclass[aps,pra,twocolumn,superscriptaddress,groupedaddress]{revtex4}

\usepackage{amssymb} \usepackage{color,graphicx} \usepackage{amsmath}
\usepackage{amsbsy} \usepackage{amsthm} \usepackage{bbm}
\usepackage{bm,bbm} \usepackage{float} \usepackage{braket}
\usepackage{placeins}
\usepackage[colorlinks=true,citecolor=blue,linkcolor=red,urlcolor=red]{hyperref}
\usepackage[sort&compress]{natbib}
\usepackage{comment}
\usepackage{stackengine}

\newcommand{\bq}{\begin{equation}} \newcommand{\eq}{\end{equation}}
\newcommand{\bqali}{\begin{equation}\begin{aligned}}
\newcommand{\eqali}{\end{aligned}\end{equation}}
\newcommand{\bqn}{\begin{equation*}}
\newcommand{\eqn}{\end{equation*}}

\newcommand\D{\operatorname{d}\!}
\renewcommand\k{{\bf k}}

\newcommand\x{{\bf x}}
\newcommand\y{{\bf y}}

\newcommand\kb{k_\text{\tiny B}}

\newcommand\acom[2]{\{#1,#2\}}

\newcommand\omegam{\omega_\text{\tiny m}}

\newcommand\gammam{\gamma_\text{\tiny m}}

\newcommand\DNSx{\mathcal S_{xx}}
\newcommand\DNSff{\mathcal S_{\text{\tiny CSL}}}

\newcommand\rC{r_\text{\tiny C}}

\graphicspath{{../Figures/}}

\begin{document}

\author{Matteo Carlesso}
\email{matteo.carlesso@ts.infn.it}
\affiliation{Department of Physics, University of Trieste, Strada Costiera 11, 34151 Trieste, Italy}
\affiliation{Istituto
Nazionale di Fisica Nucleare, Trieste Section, Via Valerio 2, 34127 Trieste,
Italy}

\author{Mauro Paternostro}
\email{m.paternostro@qub.ac.uk}
\affiliation{Centre for Theoretical Atomic, Molecular and Optical Physics, School of Mathematics and Physics, Queen\textquoteright{}s University, Belfast BT7 1NN, United Kingdom}

\title{{Opto-mechanical tests of collapse models}}

\date{\today}
\maketitle

The gap between the predictions of collapse models and those of standard quantum mechanics widens with the complexity of the involved systems. Addressing the way such gap scales with the mass or size of the system being investigated paves the way to testing the validity of the collapse theory and identifying the values of the parameters that characterize it. 

Despite increasing sensitivities which are taking experiments closer to regimes where the potential differences between collapse-based formulations and standard quantum theory should become apparent, the task of finding the precise value of the parameters of a given collapse model is nevertheless difficult. In fact, 
environmental decoherence -- which at the statistical level has the same signature as collapse models -- could mask any collapse-induced effects, thus biasing the interpretation of related experimental observations. 

The current efforts aimed at the test of collapse models can be notionally split into two broad classes: interferometric and non-interferometric tests.
The former, which aim at directly probing the validity of the quantum superposition principle, provide a natural test for any collapse model. They rely on the creation of a spatial superposition and, after a suitable time of free evolution -- necessary for the propagation of the collapse effects -- on the subsequent measurement of its interference contrast. The comparison of such contrast, which is weakened by the environmental and collapse noises, with the predictions of quantum mechanics provides experimental upper bounds to the collapse parameters.
The most successful experiments in this context have been performed using matter-wave interferometry and are extensively discussed elsewhere \cite{Toros:2017aa,Toros:2018aa}. Here we focus on the second class of experimental assessments, namely the non-interferometric ones, with the declared goal of illustrating their potential for the successful falsification of collapse models in close-to-state-of-the-art platforms. 

{The remainder of this Chapter is organised as follows: In Sec.~\ref{newper} we review the recently proposed non-interferometric approach to the testing of collapse models. Sec.~\ref{om} specialises our assessment to the opto-mechanical platform. In particular, we focus on the description of two recent thought experiments, which have paved the way to the design of experimental routes to the falsification of collapse mechanisms. In Sec.~\ref{bounds} we assess quantitative bounds provided by a set of experiments that broadly fall into the category of non-interferometric settings. Finally, Secs.~\ref{sec.cslgen} and \ref{future} address the open questions linked to plausible extensions of standard and nearly canonical formulations of collapse theories and the use of rotational degrees of freedom of mechanical rotors as ultra-sensitive tools for the inference of the minuscule effects of collapse models.}

\section{Non-interferometric experiments: a new perspective in collapse model testing}
\label{newper}
Differently from interferometric tests, where a superposition needs to be created, sustained and finally measured, non-interferometric assessments tests do not rely on the availability of high-quality non-classical resource states. 
A plethora of different experiments fall in this class, from those involving the x-ray radiation spontaneously emitted from Germanium \cite{Adler:2013aa,Bassi:2014aa,Donadi:2014aa} to those focussing on the change of the internal energy of matter-like systems~\cite{Adler:2018aa,Bahrami:2018aa,Adler:2019aa}, from the monitoring of the free expansion of cold atoms~\cite{Bilardello:2016aa} to experiments based on the dynamics of opto-mechanical systems, which are currently considered to be one of the most promising platforms for the delicate discrimination between collapse-based models and standard quantum mechanics.


\begin{figure*}[t]
\begin{center}
{\bf (a)}\hskip7cm{\bf (b)}\\
\includegraphics[width=.4\textwidth]{{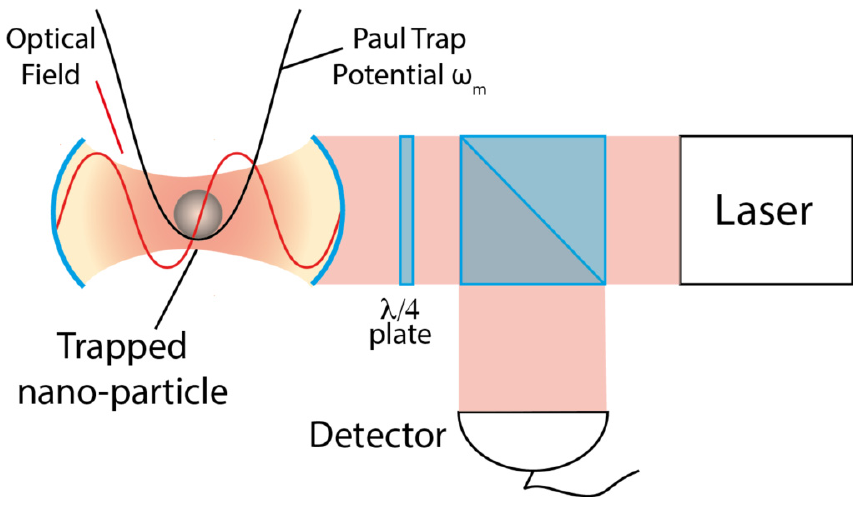}} \hspace{0.1\textwidth}\includegraphics[width=.4\textwidth]{{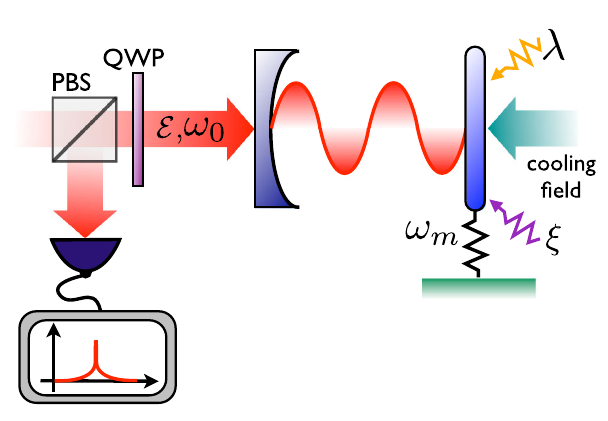}}
\end{center}
\caption{Graphical representation of two opto-mechanical setups proposed for testing collapse models. {\bf (a)}: A Paul trap, which provides the mechanism for the levitation of a charged nanoparticle, is supported by an optical cavity, required for the particle cooling. {Picture taken from Ref.~\cite{Goldwater:2016aa}}. {\bf (b)}: End-cavity opto-mechanical setup as proposed in Refs.~\cite{Bahrami:2014aa, Nimmrichter:2014aa, Diosi:2015ab}{: the cavity field is sustained by an external laser at frequency $\omega_0$. The end mirror resonates at frequency $\omegam$ and is subject to environmental noise -- described as Brownian motion at non-zero temperature and associated with the noise operator $\xi$ -- and collapse noise (described by the operator $\lambda$). {Picture taken from Ref.~\cite{Bahrami:2014aa}. }The detection scheme is the same in both the setups: a quarter-wave-plate ($\lambda/4$-plate or QWP) and a polarizing beam splitter (PBS) are used to redirect the light leaving the cavity to a detector for the reconstruction of the optical DNS.}}
\label{fig.1}
\end{figure*}

Here, we review the 
proposals put forward in Ref.~\cite{Bahrami:2014aa, Nimmrichter:2014aa, Diosi:2015ab}, which have planted the seeds for the opto-mechanical exploration of collapse models via non-interferometric approaches. For concreteness, we will focus on the Continuous Spontaneous Localization (CSL) model \cite{Ghirardi:1990aa,Pearle:1989aa,Bassi:2003aa}, which is characterized by parameters $\lambda$ and $\rC${: the first is the collapse rate, while the second is the correlation distance.}

\begin{figure*}[t!]
\centering
\includegraphics[width=0.4\linewidth]{{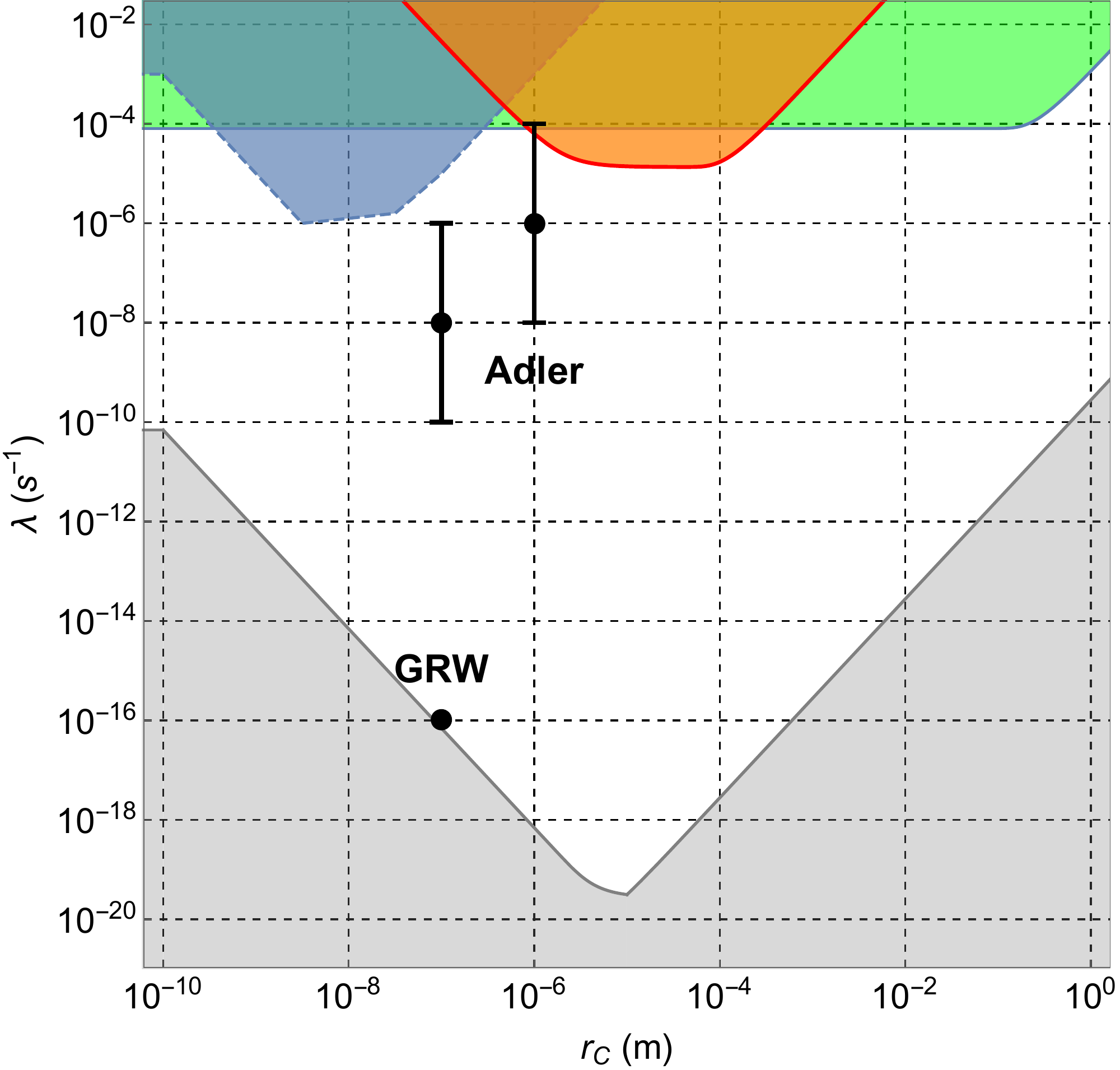}}\hspace{0.1\linewidth}\includegraphics[width=0.4\linewidth]{{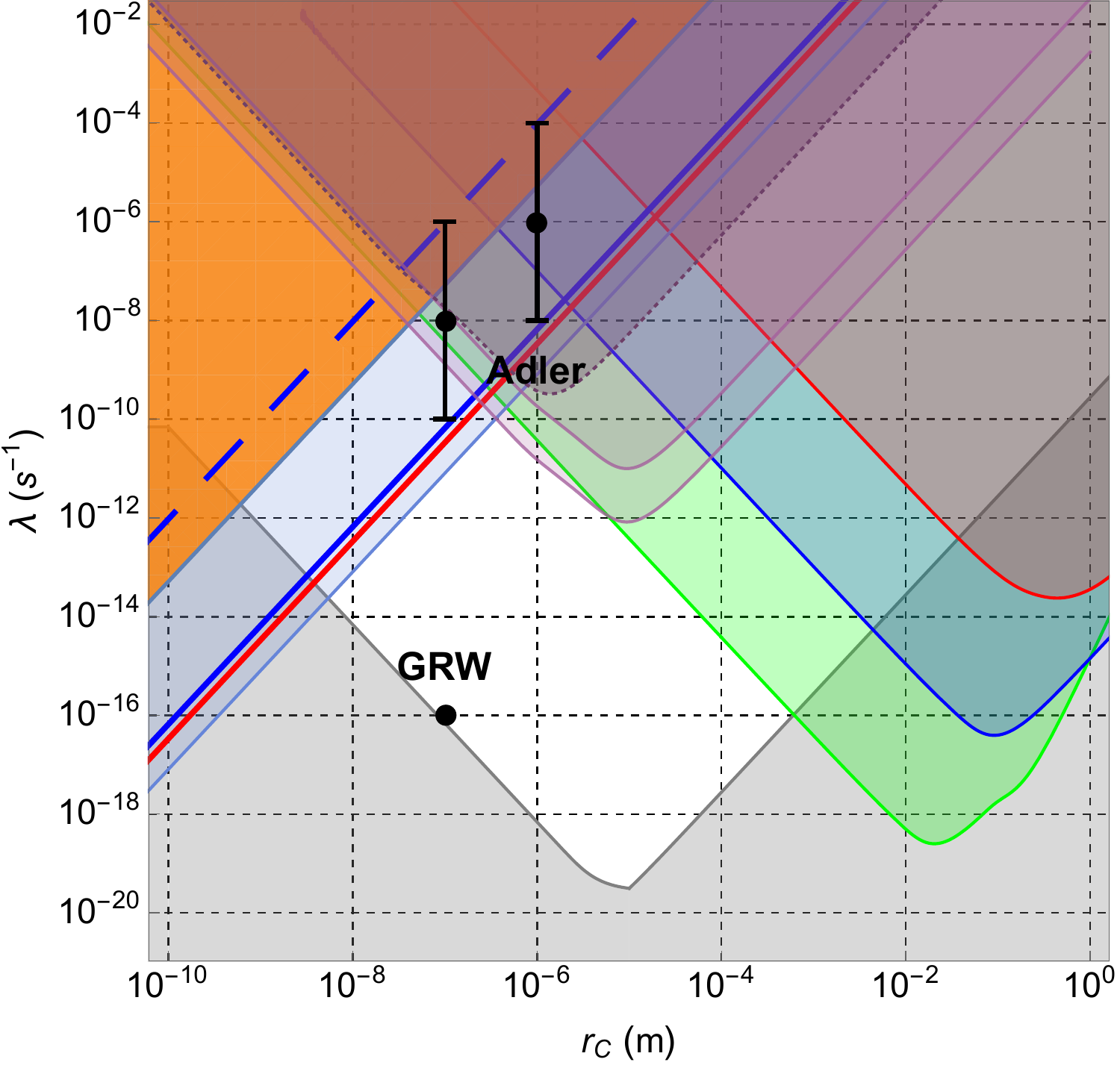}}
\caption{\label{fig.2}Exclusion plots for the CSL parameters with respect to the GRW's and Adler's theoretical values~\cite{Bassi:2003aa,Adler:2007ab}. \textit{Left panel} - Excluded regions from interferometric experiments: molecular interferometry \cite{Eibenberger:2013aa,Toros:2017aa} (blue area), atom interferometry \cite{Kovachy:2015aa} (green area) and an experiment with entangled diamonds \cite{Lee:2011aa} (orange area). {\textit{Right panel} - Regions of the parameter space of CSL excluded by a set of non-interferometric experiments: AURIGA, LIGO and LISA Pathfinder \cite{Carlesso:2016ac,Armano:2018aa} (red, blue and green areas, respectively), cold atoms \cite{Kovachy:2015ab} (orange area), phonon excitations in crystals~\cite{Adler:2018aa} (red line), blackbody radiation from the neutron star {PSR J 1840-1419} and from Neptune~\cite{Adler:2019aa} (dashed and continuous blue lines, respectively), X-ray measurements~\cite{Aalseth:1999aa,Adler:2013aa,Bassi:2014aa,Donadi:2014aa,Piscicchia:2017aa} (light blue area) and nanomechanical cantilever and its improved version~\cite{Vinante:2016aa,Vinante:2017aa} (purple areas with dashed and continuous bound). The grey color highlights the region excluded on the basis of theoretical arguments~\cite{Toros:2017aa}.}}
\end{figure*}

{To illustrate the effects induced by the CSL model, we} consider a confined system of mass $m$ whose dimensions are, for the sake of simplicity, point-like. The system is initially in thermal equilibrium at temperature $T$, which we shall assume to be small so as to make thermal fluctuations irrelevant. The confining mechanism is then switched off and the system is allowed to freely evolve for a time $t$, when measurement of the position of the system is performed. During the free evolution, the effect of the CSL mechanism can be read out in the spread of the position, which reads
\bq\label{cslx2}
\braket{\hat \x^2(t)}=\braket{\hat \x^2(t)}_\text{\tiny QM}+\frac{\lambda\hbar^2t^3}{2m_0^2\rC^2},
\eq
where $m_0$ is the mass of a nucleon, $\braket{\hat \x^2(t)}_\text{\tiny QM}$ gives the contribution due to quantum mechanics, and the last term is due to the CSL effect. There is a qualitative difference between the evolution of the spread due to quantum mechanics (which is $\sim t^2$) and the contribution arising from the collapse mechanism ($\sim t^3$). The diffusion induced by the environment has a behaviour similar to the one due the collapse mechanism~\cite{Romero-Isart:2011aa}. On this basis, a way to extrapolate the parameters of CSL would pass through the observation of the diffusive Brownian process and the consequent establishment of bounds on the collapse parameters.
This idea was put forward in Ref.~\cite{Goldwater:2016aa}, which considered a levitated charged nanosphere in a Paul trap supported by an optical cavity [the latter being needed for passive cooling of the system, cf.~Fig.~\ref{fig.1} {\bf (a)}]. Clearly, the standard decoherence sources, such as thermal photon emission, absorption and scattering as well as the collision with the residual gas particles, would also contribute to the diffusive motion of the system. The analysis performed in \cite{Goldwater:2016aa,McMillen:2017aa} is, in this context, particularly useful as reporting a comparison between possible diffusive contributions from collapse models and analogous terms resulting from standard decoherence mechanisms. By following ideas akin to those pursued in Ref.~\cite{Goldwater:2016aa}, quantitative bounds on the CSL parameters were derived from a cold atom experiment \cite{Bilardello:2016aa}, where the free expansion of the gas cloud was characterized and compared with the collapse-induced diffusion. The corresponding upper bounds are reported in Fig.~\ref{fig.2}.

\section{Opto-mechanical system as a probe of the collapse mechanism}
\label{om}

Let us now turn to the role played by opto-mechanical experiments in the assessment of collapse models. {They focus on an indirect effect provided by the collapse mechanism, which is an extra Brownian-like motion of the center of mass of the mechanical component of an opto-mechanical system. Such motion leads to an extra diffusion mechanism that can be detected through standard experimental techniques and, under suitable conditions, provide information on the undergoing collapse mechanism.} To give a concrete example, we assume a single-sided Fabry-Perot cavity endowed with an end-cavity mechanical oscillator and driven by an external laser, which also provides the mechanism for the  measurement of the mechanical motion [cf.~Fig.~\ref{fig.1} {\bf (b)}]. The latter is influenced by a phononic environment (at non-zero temperature) and, allegedly, the CSL-like collapse noise.
The action of the latter can be added to the Langevin equations governing the opto-mechanical motion, which read~\cite{Bahrami:2014aa}
\bq
\frac{\D \hat x_t}{\D t}=\frac{\hat p_t}{m},~\text{and}~
\frac{\D\hat p_t}{\D t}=-m \omegam^2\hat x_t+\hbar\chi\hat a_t^\dag\hat a_t-\gammam\hat p_t+\hat\xi_t+\hat F_t^\text{\tiny CSL},
\eq
where $\omegam$ and $\gammam$ are the harmonic frequency of the mirror and its damping constant, $\chi$ denotes the coupling of the mechanical oscillator with the cavity field, whose creation and annihilation operators are $\hat a^\dag$ and $\hat a$ respectively. Here, $\hat\xi_t$ and $\hat F_t^\text{\tiny CSL}$ denote the stochastic forces due to the environment and the collapse mechanism, respectively.
Indeed, the collapse action can be mimicked by adding to the Schr\"odinger equation a stochastic potential $\hat V_\text{\tiny CSL}$, whose corresponding force is given by $\hat F_\text{\tiny CSL}(t)=\tfrac{i}{\hbar}[\hat V_\text{\tiny CSL}(t),\hat p]$. In the case of CSL we have \cite{Adler:2013aa}
\bq\label{CSLpotential}
\hat V_\text{\tiny CSL}=-\hbar\sum_j\frac{m_j}{m_0}\int\D\x\,\hat \Psi_{ j}^\dag(\x,t)\hat \Psi_{ j}(\x,t)N(\x,t),
\eq
where $\hat \Psi_{ j}^\dag(\x,t)$ and $\hat \Psi_{ j}(\x,t)$ are respectively the creation and annihilation operators of a $j$-type particle of mass $m_j$, and $N(\x,t)$ is a stochastic noise inducing the collapse, whose mean and correlator are
\bq\label{corrnoise}
\mathbb E[N(\x,t)]=0,\ \text{and}\ 
\mathbb E[N(\x,t)N(\y,s)]=\lambda\delta(t-s)G(\x-\y),
\eq
with $\mathbb E$ the stochastic average over the noise and 
$G(\x)={e^{-\x^2/4\rC^2}}$. Eq.~\eqref{corrnoise} gives a clear interpretation of $\lambda$ and $\rC$ as the collapse rate and the noise correlation distance respectively.

The signatures of the collapses of the mechanical motion can be tracked through the density noise spectrum (DNS), whose definition reads
\bq
\DNSx(\omega)=\int\frac{\D\Omega}{4\pi}\,\mathbb E\left[\braket{\acom{\tilde x(\omega)}{\tilde x(\Omega)}}\right],
\eq
where $\tilde x(\omega)$ is the Fourier transform of the fluctuations of $\hat x_t$. Following the derivation in Ref.~\cite{Paternostro:2006aa}, one finds
\bq\label{dsnopto}
\begin{aligned}
\DNSx(\omega)&=\frac{2\hbar^2|\alpha|^2\kappa \chi^2}{m^2\left[\kappa^2+(\Delta-\omega)^2\right]|d(\omega)|^2}\\
&+\frac{	\hbar m\gammam\omega\coth\left(\tfrac{\hbar\omega}{2\kb T}\right) +\DNSff}{m^2|d(\omega)|^2},
\end{aligned}
\eq
where $|\alpha|^2$ denotes the intensity of the intra-cavity laser, $\Delta$ is the laser-cavity detuning, $T$ is the environmental temperature, and $\kappa$ is the cavity dissipation rate. Moreover we have introduced the susceptibility function $1/|d(\omega)|^2$ with
\bq
|d(\omega)|^2=(\omega_\text{\tiny m,eff}^2(\omega)-\omega^2)^2+\gamma_\text{\tiny m,eff}^2(\omega)\omega^2.
\eq
Here, $\omega_\text{\tiny m,eff}(\omega)$ and $\gamma_\text{\tiny m,eff}(\omega)$ denote the effective mechanical frequency and damping rate, respectively. Finally, $\DNSff$ quantifies the action of CSL noise, which can be obtained from $\mathbb E[\braket{\hat F_\text{\tiny CSL}(t)\hat F_\text{\tiny CSL}(t')}]=\DNSff\delta(t-t')$ with~\cite{Carlesso:2018ab}
\bq\label{dnscsl}
\DNSff=\frac{\hbar^2\lambda\rC^3}{\pi^{3/2}m_0^2}\int\D\k\,\left|\tilde \mu\left({\k}\right)\right|^2e^{-\k^2\rC^2}k_x^2,
\eq
where $\tilde \mu(\k)$ is the Fourier transform of the mass density. Here, due to the presence of the mass density, two aspects can be considered. First, $\DNSff$ is proportional to the square of the mass $m$ of the system. Thus, heavier masses can provide a stronger signature of the collapse mechanism. Second, Eq.~\eqref{dnscsl} strongly depends on the geometry of the system and in particular on the ratio between its size $L$ and $\rC$. Indeed, in the limit of $\rC\ll L$ the collapse noise will act incoherently on parts of the system which are more distant than $\rC$, while for $\rC\sim L$ such action will be coherent. Finally, for $\rC\gg L$, the collapse action will be still coherent but unfocused on the system, thus effectively loosing strength. The dependence of $\DNSff$ on the geometry of the system is clearly visible in the shape of the corresponding upper bounds on the collapse parameters. Indeed, as it is shown in Fig.~\ref{fig.1}, once the dimensions $L$ of the system are fixed, one has the strongest bound on $\lambda$ for the value of $\rC\sim L$. This reflects in the characteristic $V$-shaped form of the bounds of the CSL parameters. 

Eq.~\eqref{dsnopto} gives insight in the collapse action on the mechanical oscillator. This is the change of the equilibrium temperature of the system from the environmental one $T$ to an enhanced effective one. Indeed, in the limit for high
temperatures of the environment this reads~\cite{Carlesso:2018ab}
\bq
\hbar m\gammam\omega\coth\left(\tfrac{\hbar\omega}{2\kb T}\right) +\DNSff\to2 m\gammam\kb (T +\Delta T_\text{\tiny CSL})
\eq
with
\bq\label{tempCSL}
\Delta T_\text{\tiny CSL}=\frac{\DNSff}{2m\gammam\kb}.
\eq
One should notice that, here, another parameter of the opto-mechanical setup plays an important role, namely the damping rate $\gammam$ that quantifies mechanical dissipation. Clearly, the more the system dissipates, the faster the thermalization process to the environmental temperature, and the smaller the collapse contribution. On the contrary, in the limit of no dissipation (i.e. for $\gammam\to0$), $\Delta T_\text{\tiny CSL}$ diverges: this is exactly what should be expected from the model, whose collapse noise can be associated to an infinite-temperature bath. In passing, we remark that generalizations of collapse models have been proposed~\cite{Smirne:2015aa,Nobakht:2018aa,Smirne:2014aa} where the noise inducing collapse is associated with a finite temperature $T_\text{\tiny CSL}$ and an ensuing dissipative process. We refer to Sec.~\ref{sec.cslgen} for details on such models.

As underlined in Ref.~\cite{Nimmrichter:2014aa}, the thermal noise, proportional to $\coth(\tfrac{\hbar\omega}{2\kb T})$, is not the only limitation in detecting the collapse-induced diffusion. Indeed, the measurement process also contributes to enhancing the noise in the readout signal, thus screening the signal from the collapse mechanism. Clearly, a precise characterization of the thermal effects and the measurement backaction would provide stronger upper bounds to the collapse parameters. 

\section{Experimental bounds}
\label{bounds}

The first application that we consider 
is the one reported in Ref.~\cite{Carlesso:2016ac}, 
where three experiments -- LIGO, AURIGA and LISA Pathfinder -- have been considered. The first two are gravitational wave detectors, while the last one is only a prototype of a future gravitational wave detector. In all such experiments, a mechanical resonator is monitored through optical techniques. Due to the mass of the systems ($\sim2$\,kg for LISA Pathfinder, $\sim40\,$kg for LIGO and $\sim2300\,$kg for AURIGA), the back-action of the optics can be neglected, and one considers only the last term in Eq.~\eqref{dsnopto}, which depends explicitly on the experiment considered. The single arm of LIGO and LISA Pathfinder consists of two masses, modelled as harmonic oscillators, whose relative distance is monitored. Conversely, AURIGA is a resonant bar whose elongation is measured. For the latter, one can model the system as two half-mass harmonic oscillators oscillating in counterphase. Thus, the modelling is the same for all three experiments. Eq.~\eqref{dnscsl} is consequently modified to read
\bq\label{dnscslrel}
\DNSff=\frac{\hbar^2\lambda\rC^3}{2\pi^{3/2}m_0^2}\int\D\k\,\left|\tilde \mu\left({\k}\right)\right|^2e^{-\k^2\rC^2}k_x^2(1-e^{i a k_x}),
\eq
where $a$ is the distance between the two masses.
Such systems are well outside the quantum realm due to their masses, which also prevent their use in interferometric experiments. However, they set important bounds on the collapse parameters, which are here reported in Fig.~\ref{fig.1}.

The second application that we aim at covering is that reported in Refs.~\cite{Vinante:2016aa,Vinante:2017aa}, where a heavy micrometrical sphere is attached to a silicon cantilever, which acts as a mechanical resonator. As the sphere is ferromagnetic, in place of the optics, a low noise SQUID can be employed to monitor the mechanical motion of the cantilever. The system is placed in high vacuum and low temperature to minimize the thermal action of the environment. Moreover, in order to better characterize the thermal component of the noise, different measurements of the DNS of the system were performed at different temperatures of the environment, ranging from $11$\,mK to $\sim1\,$K. Thus, by exploiting Eq.~\eqref{tempCSL}, one can determine upper bounds on the collapse parameters $\lambda$ and $\rC$, which are reported in Fig.~\ref{fig.1}.

\section{Testing of the dissipative and colored CSL models}\label{sec.cslgen}

\begin{figure*}[t!]
\centering
\includegraphics[width=0.25\linewidth]{{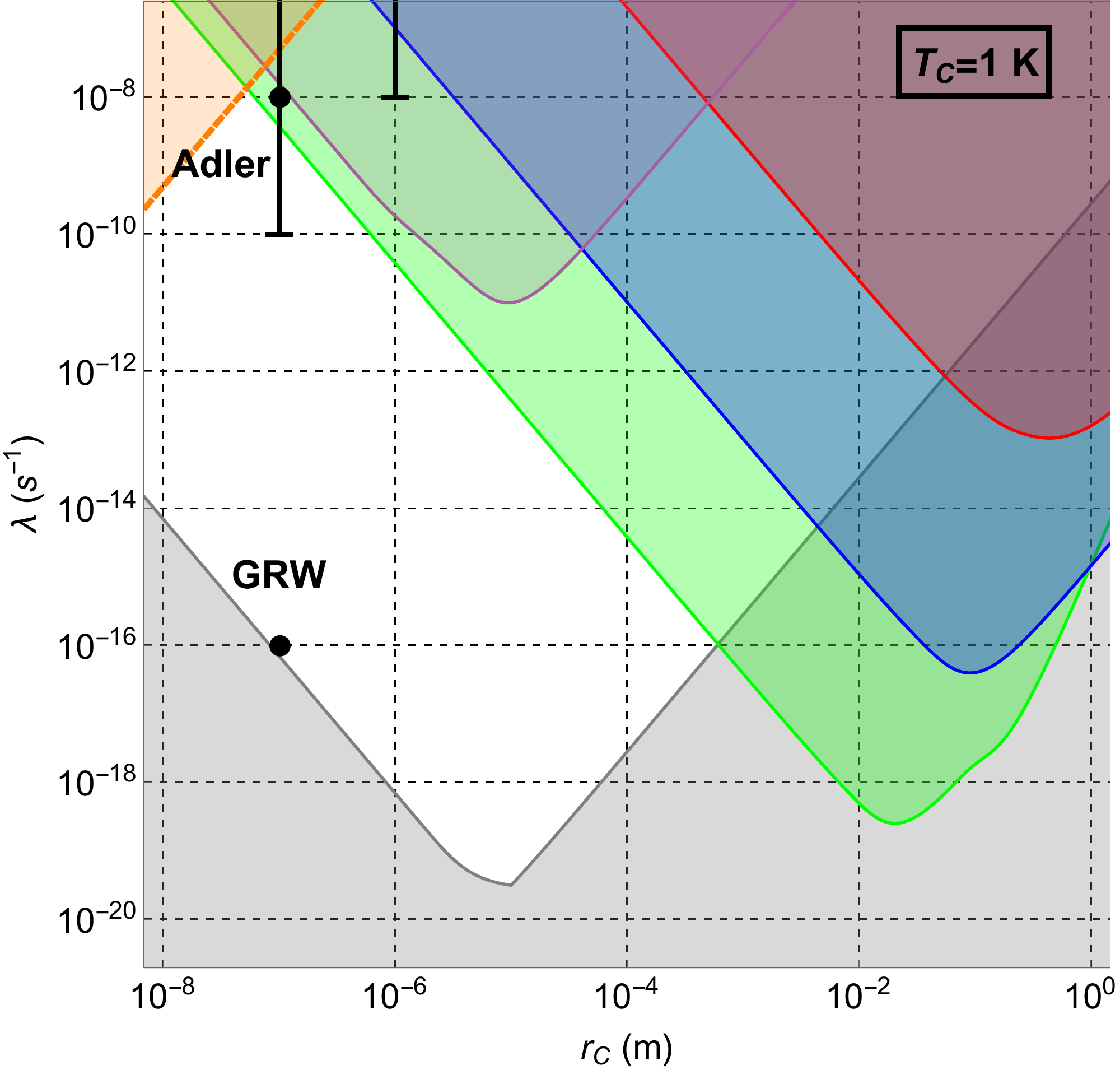}}\includegraphics[width=0.25\linewidth]{{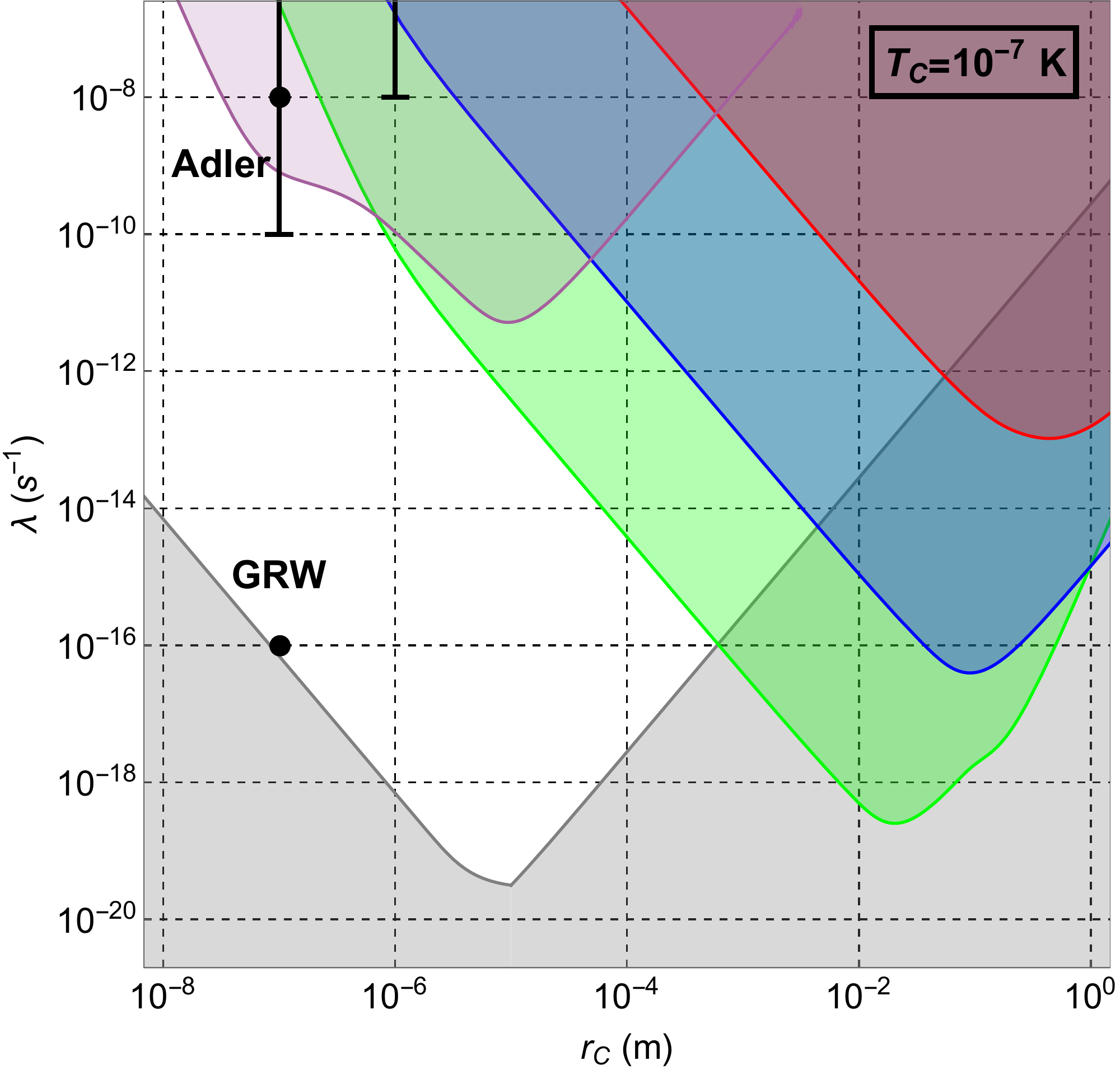}}\includegraphics[width=0.25\linewidth]{{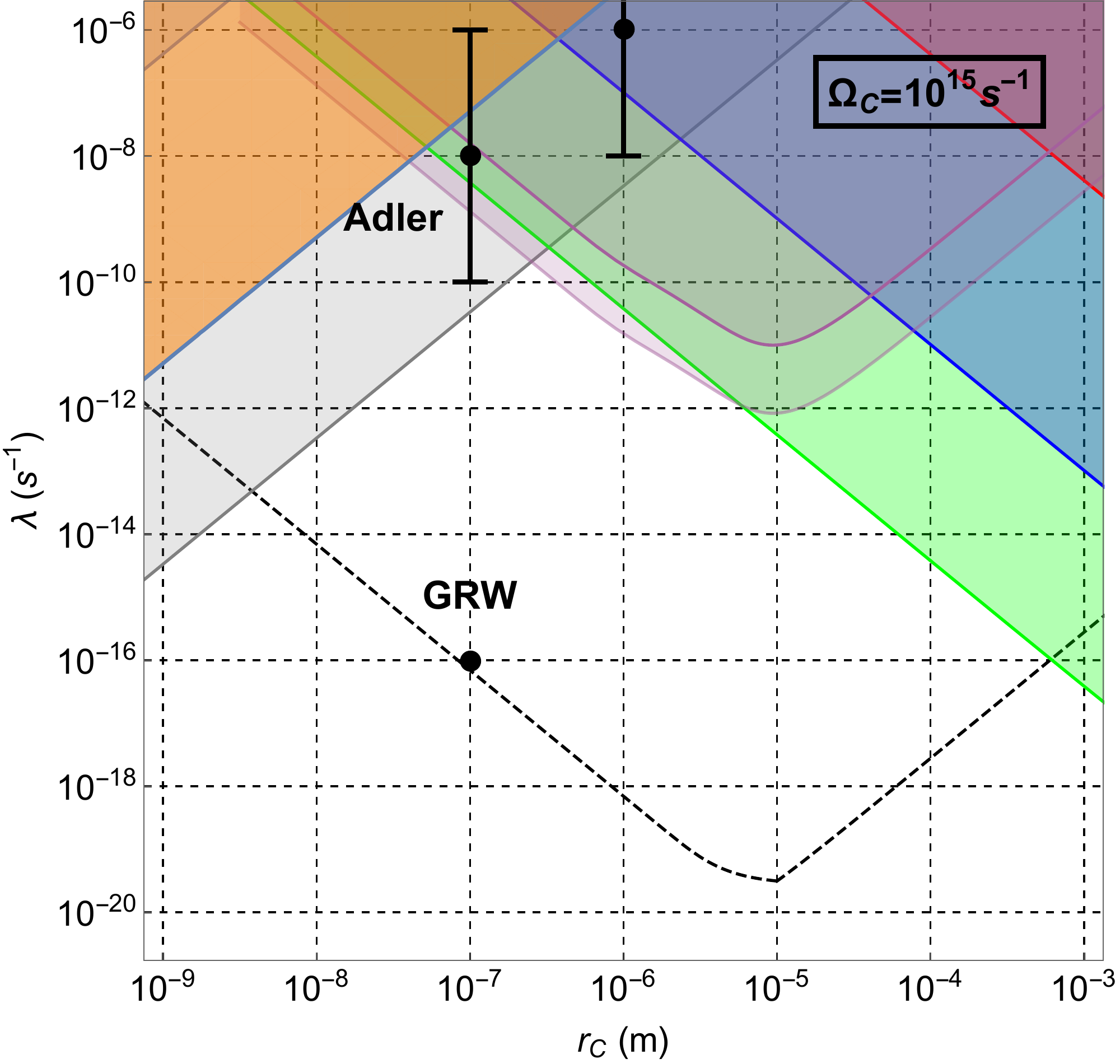}}\includegraphics[width=0.25\linewidth]{{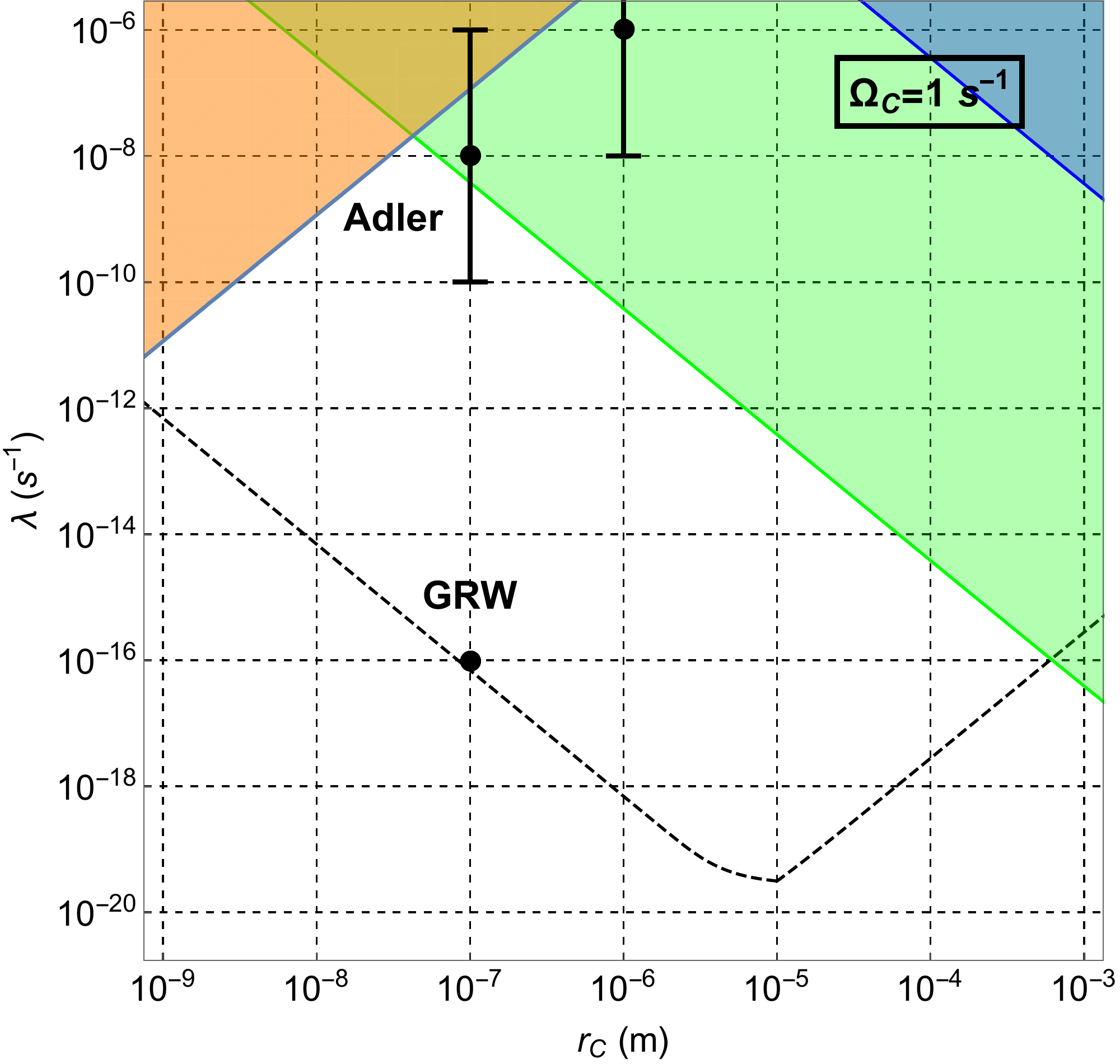}}
\caption{\label{fig:extensions} \textbf{First and second panels}: Upper bounds on the dissipative CSL parameters $\lambda$ and $\rC$ for {two} values of the CSL noise temperature: $T_\text{\tiny CSL}=1\,$K (first panel) and $T_\text{\tiny CSL}=10^{-7}\,$K (second panel). {Picture taken from \cite{Nobakht:2018aa}}. \textbf{Third and fourth panels}: Upper bounds on the colored CSL parameters $\lambda$ and $\rC$ for two values of the frequency cutoff: $\Omega_\text{\tiny c}=10^{15}\,$ Hz (third panel) and $\Omega_\text{\tiny c}=1\,$ Hz (fourth panel). {Picture taken from \cite{Carlesso:2018aa}.}
Red, blue and green lines (and respective shaded regions): Upper bounds (and exclusion regions) from AURIGA, LIGO and LISA Pathfinder, respectively \cite{Carlesso:2016ac}. Purple region: Upper bound from cantilever experiment \cite{Vinante:2017aa}.  Orange and grey top regions: Upper bound from cold atom experiment \cite{Kovachy:2015ab,Bilardello:2016aa} and  from bulk heating experiments \cite{Adler:2018aa}.
The bottom area shows the excluded region based on theoretical arguments \cite{Toros:2017aa}. }
\end{figure*}

The CSL model have two weaknesses~\cite{Bassi:2003aa}. The first is the steady increase in the energy of any (free) system in time, e.g.~an hydrogen atom is heated by $\simeq 10^{-14}$\,K per year taking the values $\lambda=10^{-16}$\,s$^{-1}$ and $\rC=10^{-7}$\,m. Although the increment is small, it is not realistic feature even for a phenomenological model. On the other hand, one expects that, through a dissipative mechanisms, the system will eventually termalize to the finite temperature of the collapse noise. Although there are theoretical arguments suggesting the value of such a temperature to be $T_\text{\tiny CSL}\simeq1\,$K \cite{Smirne:2015aa,Smirne:2014aa}, one needs to validate them. 
While an interferometric investigation was performed in Ref.~\cite{Toros:2018aa,Toros:2017aa}, and a non-interferometric measurement of the free expansion of a cold-atom cloud was studied in Ref.~\cite{Bilardello:2016aa},
the theoretical setting for an opto-mechanical test of the dissipative extension of the CSL model was proposed in Ref.~\cite{Nobakht:2018aa}. Fig.~\ref{fig:extensions} shows how the experimental bounds change when the dissipation is explicitly considered in the collapse mechanism for two values of the $T_\text{\tiny CSL}$.

The second weakness of the CSL model is that its noise has a white spectrum. This is clearly an approximation as no physical noise can be perfectly white. Conversely, one expects the existence of a cutoff frequency $\Omega_\text{\tiny C}$ above which the collapse mechanism is negligible. Theoretical arguments suggest $\Omega_\text{\tiny C}\sim10^{12}\,$Hz \cite{Adler:2007aa,Adler:2008aa}.  The introduction of the cutoff changes the predictions of the model: the correlations of the noise in Eq.~\eqref{corrnoise} are modified in $\mathbb E[N(\x,t)N(\y,s)]=\lambda f(t-s)G(\x-\y)$, where $f(t)$ describes the time correlations of the collapse noise. Correspondingly, the DNS in an opto-mechanical system becomes $\mathcal S_\text{\tiny cCSL}(\omega)=\DNSff\times\tilde f(\omega)$, where $\tilde f(\omega)$ is the Fourier transform of $f(t)$ \cite{Carlesso:2018aa}.
Bounds on the CSL parameters for colored noise were studied in detail in Ref.~\cite{Bilardello:2016aa,Toros:2017aa, Carlesso:2018aa}. In particular, upper bounds from high frequencies experiments (or involving small time scales) are weakened when moving to small value of $\Omega_\text{\tiny C}$.  Fig.~\ref{fig:extensions} shows the upper bounds to the colored CSL extension for two values of $\Omega_\text{\tiny C}$.

\section{Proposals for future testing}\label{prop}
\label{future}

\begin{figure*}[t!]
\centering
\includegraphics[width=0.25\linewidth]{{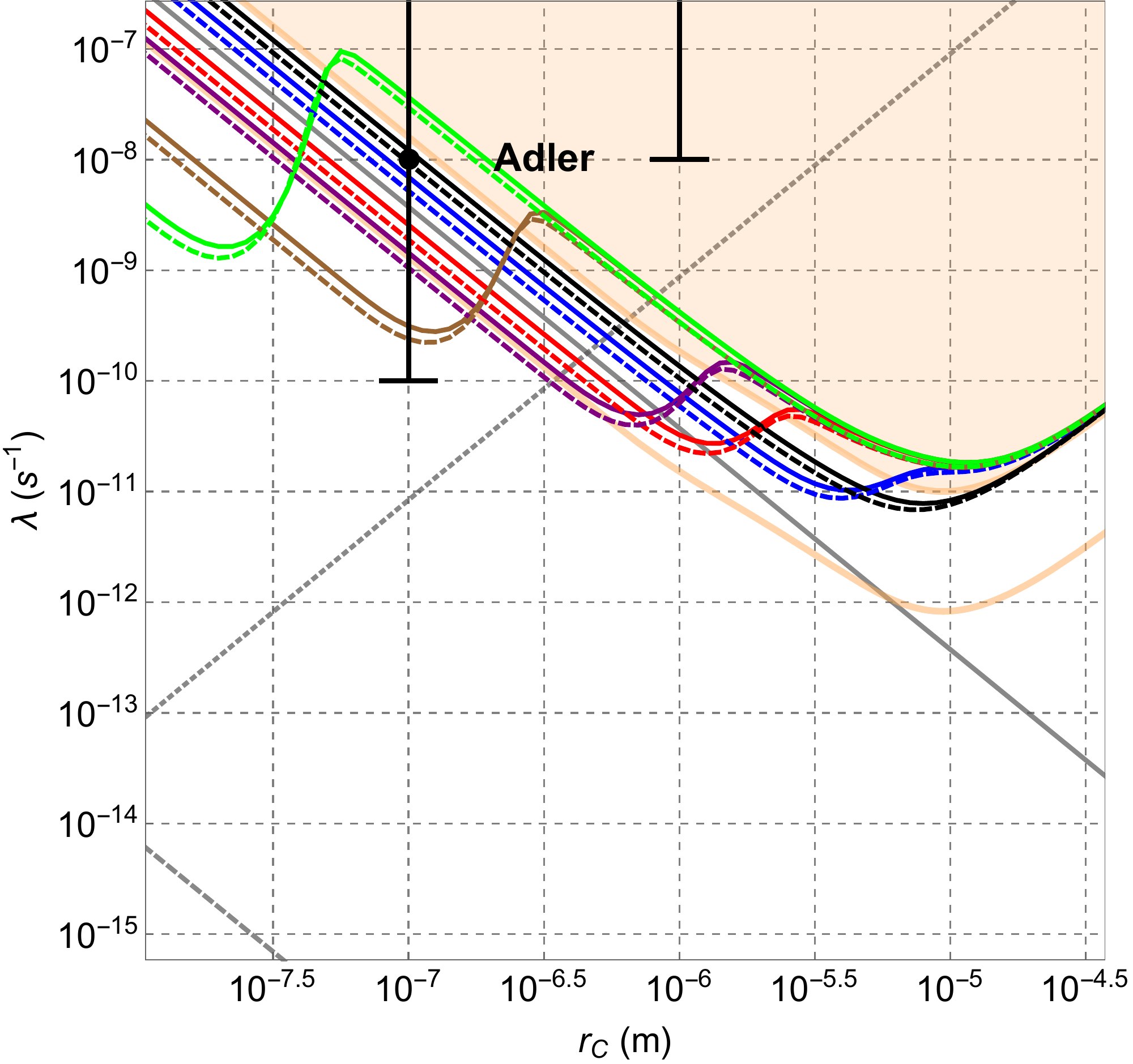}}\includegraphics[width=0.25\linewidth]{{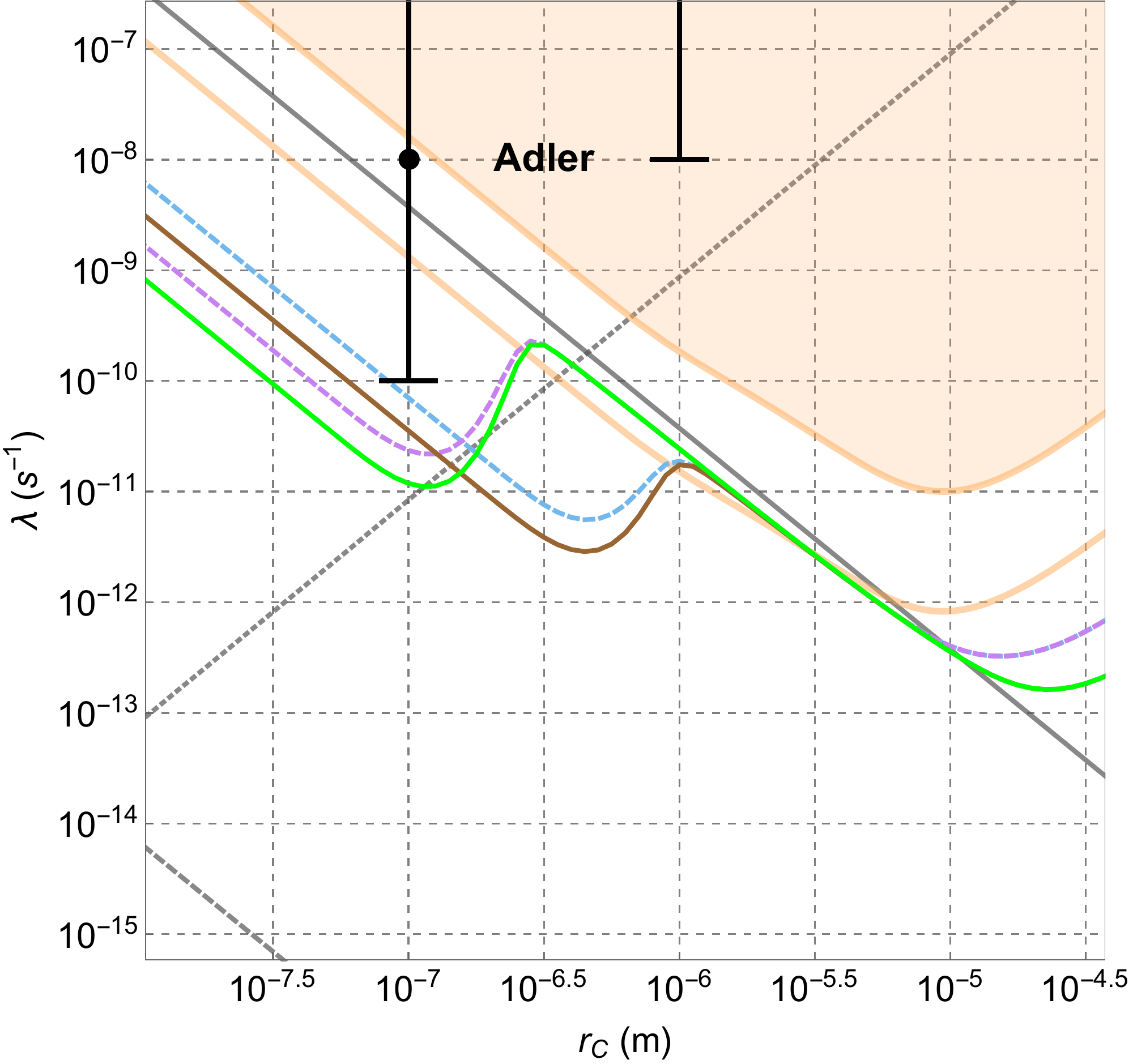}}\includegraphics[width=0.25\linewidth]{{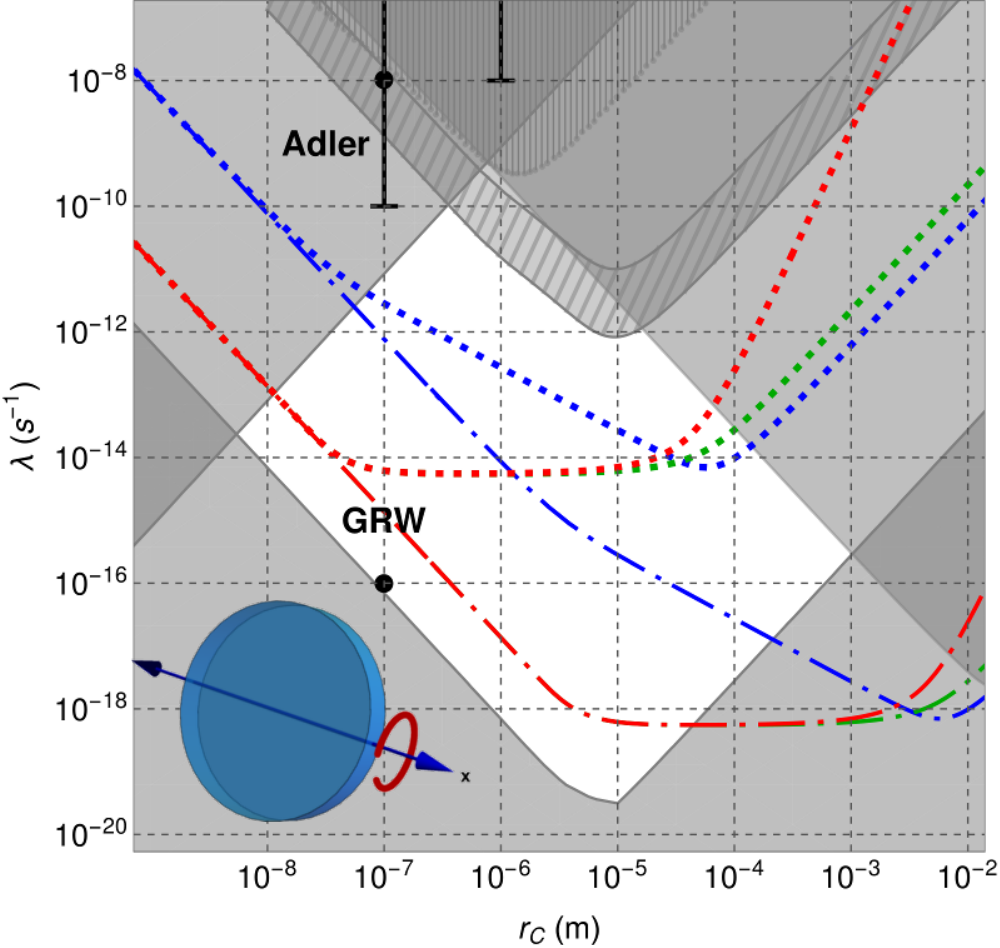}}\includegraphics[width=0.25\linewidth]{{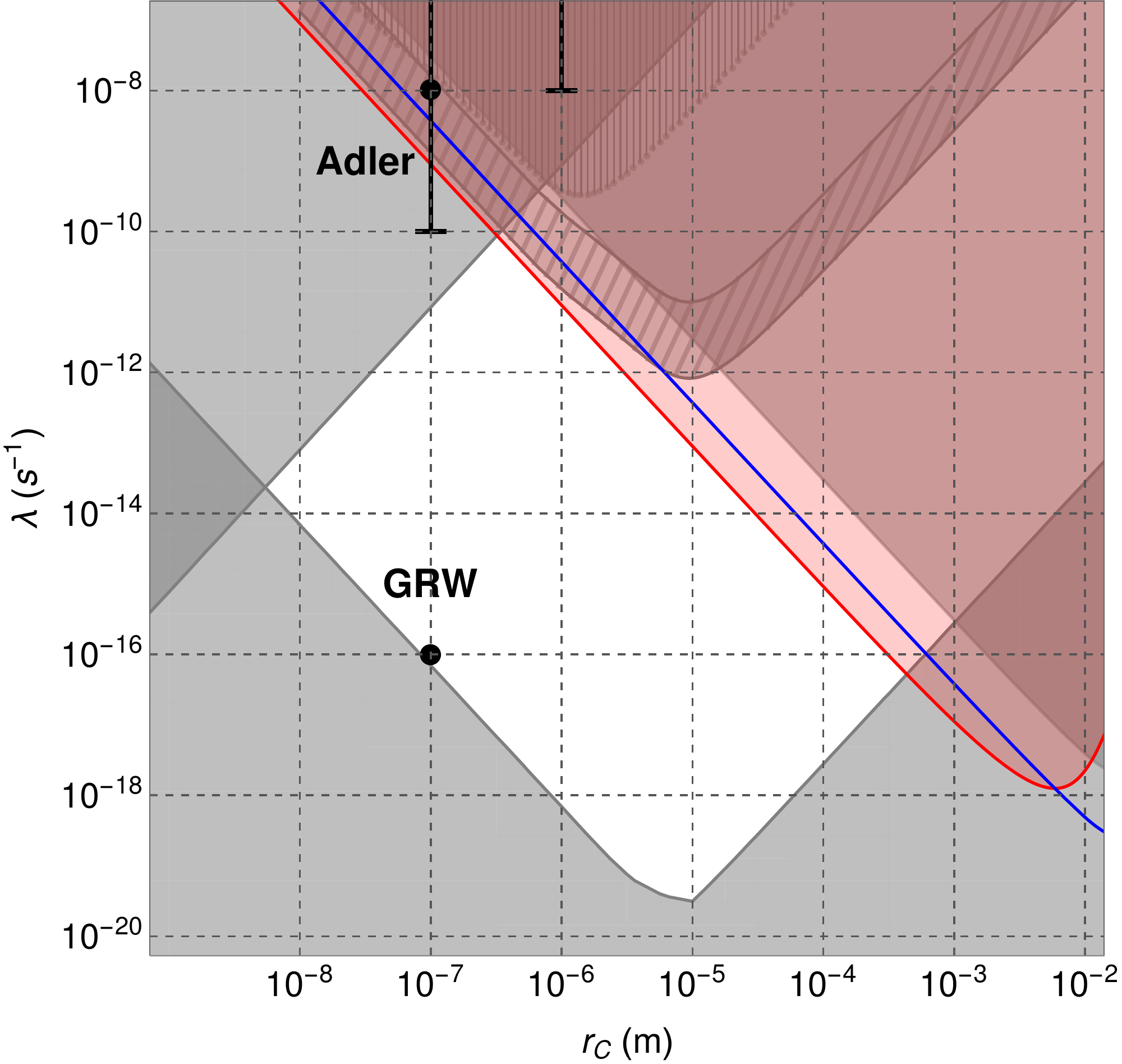}}
\caption{\label{fig:proposals} Exemplification of two possible experimental tests of collapse models. \textbf{First panels}: Hypothetical upper bounds obtained from substituting the sphere attached to the cantilever used in \cite{Vinante:2017aa} with a multilayer cuboid of the same mass for various thickness of the layers \cite{Carlesso:2018ac}. The bounds are compared with that from the improved cantilever experiment \cite{Vinante:2017aa} shown in orange. {Picture taken from \cite{Carlesso:2018ac}.} \textbf{Second panel}: Same as the first panel, but with a mass ten times larger. {Picture taken from \cite{Carlesso:2018ac}.} \textbf{Third panel}: Results of the analysis proposed in \cite{Schrinski:2017aa,Carlesso:2018ab} where the rotational degrees of freedom of a cylinder are studied. The red line denotes the upper bound that can be obtained from the constrains given by the rotational motion, compared with those from the translations (blue and green lines). {Picture taken from \cite{Carlesso:2018ab}.} \textbf{Fourth panel}: Red shaded area highlights the hypothetical excluded value of the collapse parameters that could be derived from the conversion of the translational noise of LISA Pathfinder to rotational one \cite{Carlesso:2018ab}. This is compared to the new (old) upper bounds from the translational motion shown with the blue line \cite{Carlesso:2018ab} (grey area \cite{Carlesso:2016ac}). {Picture taken from \cite{Carlesso:2018ab}}.}
\end{figure*}

Opto-mechanical proposals have been put forward aimed at strengthening the current upper bounds on the collapse parameters. One consists in the modification of the cantilever experiment in Ref.~\cite{Vinante:2017aa}, where the homogeneous mass is substituted with one made of several layers of two different materials \cite{Carlesso:2018ac}. This will increment the effect of the CSL noise for the values of $\rC$ of the order of the thickness of the layers. The hypothetical upper bounds that can be inferred from such scheme are shown in Fig.~\ref{fig:proposals}.

A second possible test focuses on the rotational degrees of freedom in place of the vibrational ones \cite{Schrinski:2017aa,Carlesso:2018ab}. The former can quantify the CSL action in a form similar to that in Eq.~\eqref{tempCSL}, where the collapse-induced contribution to the temperature is related to the rotational degrees of freedom and reads $\Delta T_\text{\tiny CSL}^\text{\tiny rot}={\DNSff^\text{\tiny rot}}/{2\kb D_\phi}$, where $D_\phi$ is the rotational damping rate and
\bq
\label{roto}
\DNSff^\text{\tiny rot}=\frac{\hbar^2\lambda\rC^3}{\pi^{3/2}m_0^2}\int\D\k\,\left|	k_y\partial_{k_z}\tilde \mu(\k)-k_z\partial_{k_y}\tilde \mu(\k)\right|^2e^{-\rC^2\k^2}.
\eq
Eq.~\eqref{roto} quantifies the stochastic torque induced on the system by the collapse noise. When such a scheme is applied to macroscopic systems, it can provide a sensible improvement of the bounds on the collapse parameters, cf.~Fig.~\ref{fig:proposals}. A direct application was considered in \cite{Carlesso:2018ab}, where the bound from LISA Pathfinder \cite{Carlesso:2016ac} can be significantly improved by also considering the rotational degrees of freedom.

The above are only two of several proposals~\cite{Collett:2003aa,Goldwater:2016aa,Kaltenbaek:2016aa,McMillen:2017aa,Mishra:2018aa} suggested over the past few years aimed to push further the exploration of the CSL parameter space.

\acknowledgments

The authors acknowledge the invaluable collaboration with A. Bassi, H. Ulbricht, S. Donadi, A. Vinante on the topics of this work. They are grateful acknowledge support from the H2020 Collaborative project TEQ (grant agreement 766900) and COST Action CA15220. MC acknowledges support from INFN. MP acknowledges support from the DfE-SFI Investigator Programme (grant 15/IA/2864), the Royal Society Wolfson Research Fellowship (RSWF\textbackslash R3\textbackslash183013) and the Leverhulme Trust Research Project Grant (grant nr.~RGP-2018-266).

%


\end{document}